\documentclass[showpacs,nofootinbib,preprint,11 pt]{article}
\usepackage{amsmath,amsfonts,float,graphicx,color,xcolor,epsfig,mathrsfs}
\usepackage{epstopdf}
\usepackage{hyperref}
\usepackage{slashed,comment}
\usepackage{footnote}
\usepackage{caption, subcaption}
\usepackage[bottom]{footmisc}
\usepackage[normalem]{ulem}
\usepackage[pass]{geometry}
\begin{document}
\newgeometry{left=2.54cm,top=2cm,bottom=2cm,right=2.54cm}
\begin{center}
\setlength {\baselineskip}{0.3in} 
\def \pt{\partial}
\def \L{\mathcal{L}}
\def \O{\mathcal{O}}
\def \D{\mathcal{D}}
\def \Tr{\textmd{Tr}}
\def \STr{\textmd{STr}}
\def \and{\textmd{and}}
\def \nn{\nonumber} 
\def \TT{\textmd{TT}} 
\def \T{\textmd{T}} 

\def \be{\begin{equation}}
\def \ee{\end{equation}}

\def \bea{\begin{eqnarray}}
\def \eea{\end{eqnarray}}
{\bf\Large\boldmath 
Radiative $B$ to tensor meson decays at NLO in SCET
}\\[5mm]

\setlength {\baselineskip}{0.2in}
\setlength {\baselineskip}{0.3in}
{\large Arslan Sikandar{\footnote{Current Address: Rudjer Boskovic Institute, Division of Theoretical Physics, Bijenička 54, HR-10000 Zagreb, Croatia.} \footnote{Corresponding Author}}, M. Jamil Aslam\\
			
Department of Physics, Quaid-i-Azam University, Islamabad 45320, Pakistan.}
\end{center}

{\bf Abstract}\\[5mm] 
\setlength{\baselineskip}{0.2in} 

The radiative $B$ to tensor $\left(K_2^*(1430),\; f_2(1270),\; a_2(1320)\right)$ meson decays are studied at next-to-leading order (NLO) in soft-collinear effective theory (SCET). The SCET allows the systematic treatment of factorizable and non-factorizable contributions along with the resummation of large perturbative logarithms. We performed a two step matching and determined the soft-overlap function $\zeta^\perp_{T}$ and branching ratios for these $B$ to tensor meson decays. In the case of $B\to K_2^*(1430)\gamma$, the numerical value of the branching ratio lies close to its experimental measurements. The estimated values of the branching ratios of CKM suppressed decays $B \to \left(a_2(1320),\;  f_2(1270)\right)\gamma$ are significant small compared to that of  $B\to K_2^*(1430)\gamma$, but still could be measured in some ongoing and future $B$ physics experiments. 

%\usepackage{palatino}

%-------------------------------------------------------
\section{Introduction}
The weak $B$ to pseudosclar $(P)$, vector $(V)$, axial-vector $(A)$ meson decays involving the quark level flavor changing neutral current transitions (FCNC) transitions $b\to (s,d)$, occur at loop level in the Standard Model (SM); therefore, these are suppressed both by the loop factors and the elements of Cabibbo-Kobayashi-Maskawa (CKM) matrix. This makes these FCNC transitions a fertile ground to hunt for the new physics (NP), i.e., physics beyond the SM (BSM). The radiative heavy-to-light, $B\rightarrow V\gamma$ decays are thus studied extensively, both within SM \cite{Misiak:2015xwa,Ball:2006eu,Faustov:1992xv,Khodjamirian:2010vf}
and in BSM approaches \cite{Oshimo:1992zd,Jung:2012vu}. Similar decay involving axial-vector meson in the final state, i.e., $B\rightarrow A\gamma$ is calculated in context of large-energy-effective theory (LEET) \cite{Sikandar:2019qyb} and using soft-collinear effective theory (SCET) \cite{Sikandar:2021shv}. The radiative $B\to K_2^*(1430)$, where $K_2^*(1430)$ is a tensor mesons, is studied using perturbative QCD approach \cite{Wang:2010ni}, light-cone sum rules \cite{Yang:2010qd,Aliev:2019ojc}, Light front quark model (LFQM) \cite{Cheng:2003sm,Cheng:2009ms}, Heavy Quark Effective Theory (HQET) \cite{Veseli:1995bt}, and LEET \cite{Sikandar:2022iqc}. Also, the experimental averaged value of the branching ratio of $B\to K_2^*(1430)\gamma$ is measured to be $(1.24\pm 0.24) \times 10^{-5}$ \cite{Workman:2022ynf}.
 
This study focuses the radiative $B$ to tensor meson $(J^P=2^+)$ decays using the framework of SCET. The tensor mesons $K_2^*(1420), a_2(1320), f_2(1270)$ are lighter compared to $B$ meson, and this justifies the use of  SCET for the underlying $b\rightarrow (s,d)\gamma$ transition, where the masses of $s$ and $d-$quarks are quite small compared to $b-$quark. The advantage of SCET is the ease it provides to study next-to-leading-order (NLO), i.e.,  $(\mathcal{O}(\alpha_s))$ corrections, power corrections and resummation. Moreover, SCET allows to divide systematically the heavy-to-light decay in terms of factorizable and non-factorizable contributions.

The SCET factorization for heavy-to-light decays at leading power in $1/m_b$ that is valid at all orders of $\mathcal{O}(\alpha_s)$ is 
\begin{equation}
\langle T \gamma|Q_i|B\rangle = C_i ^{I}\zeta_{B\rightarrow T}+\int_0 ^\infty \frac{d\omega}{\omega}\phi_B(\omega)\int_0 ^1 du\  \phi_{T}(u)C_i ^{II}(\omega,u),\label{theorem}
\end{equation}
where $C_i^I$ and $C_i^{II}$ are Wilson Coefficients (WCs) containing the perturbative dynamics at hard scale $(\mu_Q\sim m_b)$ and intermediate scale $(\mu_i\sim \sqrt{m_b\Lambda_{\text{QCD}}})$, respectively. $\phi_T$ and $\phi_B$ are light-cone distribution amplitude (LCDAs) of tensor and $B$ mesons, respectively. $\zeta_{B\rightarrow T}$ is the soft form-factor, also called the soft-overlapping function at large recoil $(q^2=0)$, accounts for the non-factorizable dynamics. The factorization Eq. (\ref{theorem}) is required to correctly account the factorizable corrections in WCs \cite{Bauer:2001rd,Beneke:2003pa,Charles:1998dr,Lange:2003pk}. 
Using the SCET formalism with the factorization relation Eq. (\ref{theorem}), we have to perform a two step matching procedure. In the first step QCD is matched onto SCET$_{\text{I}}$ at hard scale $\mu_Q$ and later SCET$_{\text{I}}$ is matched onto SCET$_{\text{II}}$ at intermediate scale $\mu_i$. This would require to introduce operators for the two effective theories using a power counting scheme. The introduction of these SCET operators make the resummation of the large logarithms possible. During this process, the resummation of $C_i^I$ is avoided because of the hard scale, however, for the $C_i^{II}$ it is performed due to lower value of the intermediate scale (for details: see \cite{Sikandar:2021shv}). 

This scheme is motivated from the pioneering work done by Becher \textit{et. al.} \cite{Becher:2005fg} for $B\to V\gamma$ decays, where the SCET treatment of this radiative decay at NLO in $\alpha_s$ and at leading power of $1/m_b$ has been done pedagogically. After introducing the relevant SCET operators, and taking the results of SCET$_{\text{I}}$ anomalous dimensions from \cite{Hill:2004if}, the necessary matching calculations are performed. Later, using the calculation of SCET form factor for $B\to K^*\gamma$, its branching ratio is estimated \cite{Becher:2005fg}. Ref. \cite{Sikandar:2021shv} discussed the similar calculations for the radiative $B\to A\gamma$, where $A$ is representing an axial-vector meson. In \cite{Ali:2006ew}, the authors carried a study for the semileptonic $B \to K^*$ decay, i.e., $B\to K^*\ell^+\ell^-$ at NLO in $\alpha_s$ and leading power in $1/m_b$ expansion. Constraining the soft form factor, i.e., $\zeta_{\perp}\left(q^2=0\right)$ from the radiative  $B\to K^*\gamma$ data, and using this together with the light-cone sum rules to determine the $q^2$-dependence of $\zeta_{\perp}\left(q^2\right)$ the value of branching ratio is estimated in $q^2=\left[1,\; 7\right] \text{GeV}$ range. Compared to the branching ratio of the semi-leptonic $B-$ meson decays, the zero-position of the forward-backward asymmetry is less sensitive to different input parameters. In SM, its value is estimated to be $q_0^2 = \left(4.07^{+0.16}_{-0.13}\right)$  \cite{Ali:2006ew}.

The work is organized as follows: In Sect. \ref{sec1}, we define the relevant operators for $B\to T\gamma$ decay and discuss the momentum scaling. In Sect. \ref{sec2}, we work out the matching calculation from QCD to the SCET$_{\text{I}}$. In Sect. \ref{sec4}, the SCET$_{\text{II}}$ operators are given as four-quark operators to match with the SCET$_{\text{I}}$ operators and the relevant WCs, known as the jet-functions, are presented. These WCs are scale dependent and their evolution from hard to an intermediate scale involves the resummation, which we discuss in Sect. \ref{sec5}. We calculate the matrix elements for $B\rightarrow T\gamma$ in Sect. \ref{sec6}, and present the results for the branching ratios and compare them with some earlier results and the experimental measurements. Finally,  Sect. \ref{sec7} concludes of our main results. 

\section{SCET analysis of $B\rightarrow (K_2,f_2,a_2)\gamma$}\label{sec1}

In SM, the weak effective Hamiltonian (WEH) for the FCNC, $b\rightarrow (s,d)\gamma$, transitions is given by 
\begin{equation}
\mathcal{H}_W = \frac{G_F}{\sqrt{2}}\sum_{p^\prime=u,c} V^*_{p^\prime p}V_{p^\prime b} \sum_{i=1}^8\left[C_i(\mu) Q_i(\mu)\right],\label{Heff-1}
\end{equation}
where $p=s$ for $K_2^*$ and $p=d$ for $(f_2,a_2)$ mesons. Utilizing the unitary relation and ignoring the loop contribution $V^*_{us(d)}V_{ub}$, which is doubly Cabibbo suppressed, we can write 
\begin{equation}
\mathcal{H}_W = -\frac{G_F}{\sqrt{2}}\sum_{i=1}^8 V^*_{tp}V_{tb}\left[C_i(\mu) Q_i(\mu)\right].\label{Heff}
\end{equation}
 The WCs are known at next-to-next leading logarithm (NNLL) \cite{Bobeth:1999mk, Gorbahn:2005sa, Czakon:2006ss} while the four-quark operators are \cite{Blake:2016olu}:
\begin{eqnarray}
Q_1 &=& (\bar{p}_L\gamma_\mu T^a c_L)(\bar{c}_L \gamma^\mu T^ab_L),\qquad\qquad\quad Q_2 = (\bar{p}_L\gamma_\mu c_L)(\bar{c}_L \gamma^\mu b_L),\notag\\
Q_3 &=& (\bar{p}_L\gamma_\mu b_L)\sum_q(\bar{q} \gamma^\mu q),\qquad\qquad\qquad\quad Q_4 = (\bar{p}_L\gamma_\mu T^a b_L)\sum_q(\bar{q} \gamma^\mu T^a q),\notag\\
Q_5 &=& (\bar{p}_L\gamma_\mu\gamma_\nu\gamma_\sigma b_L)\sum_q(\bar{q} \gamma^\mu\gamma^\nu\gamma^\sigma q),\qquad\quad Q_6 = (\bar{p}_L\gamma_\mu\gamma_\nu\gamma_\sigma T^a b_L)\sum_q(\bar{q} \gamma^\mu\gamma^\nu\gamma^\sigma  T^a q),\notag\\
Q_7&=&\frac{e}{16\pi^2}m_b (\bar{p}_L \sigma^{\mu\nu}b_R)F_{\mu\nu}, \qquad\qquad\qquad Q_8=\frac{g_s}{16\pi^2}m_b (\bar{p}_L \sigma^{\mu\nu}T^a b_R)G_{\mu\nu}^a.\label{EWH}
\end{eqnarray}
 In above equation, $T^a$ is the $SU(3)$ color matrix, and $m_b$ is the $b-$quark mass in $\overline{\text{MS}}$ scheme at momentum scale $\mu$. The operator $Q_2$ contribute at $\alpha_s^2$, therefore, it is ignored for these NLO calculations. Also, the WCs for penguin operators $Q_3-Q_6$ are too small, and hence we do not include their contributions in this work. The phenomenologically relevant operators for the radiative $B\rightarrow T\gamma$ decays at $\mathcal{O}(\alpha_s)$ are $Q_1 ,Q_7$ and $Q_8$. Since the masses of these tensor mesons are around $1.5$ GeV, which is about $3.5$ times smaller than the mass of the $B$ meson, they have a high collinear momentum at maximum recoil in these radiative decays. We assume that the outgoing meson is moving in a light-like direction with an associated vector $n_\mu$. The velocity vector associated with the decaying $B$ meson is $v_\mu$, which is related to $n_\mu$ through an auxiliary light-like vector $\bar{n}_\mu$ as
\begin{equation}
\bar{n}^\mu
= 2v^\mu-n^\mu ,
\end{equation}
satisfying $n^2=\bar{n}^2=0$, $n\cdot \bar{n}=2$, and correspondingly $v\cdot v=1$. In $B \to T\gamma$ decay, the momenta of $B,\; T$, and $\gamma$ are
\begin{eqnarray}
p_B^\mu =m_B v^\mu,\qquad p_T^\mu&=En^\mu +\frac{m_T^2}{4E}\bar{n}^\mu, \qquad p_\gamma^\mu = \left(E-\frac{m_T^2}{4E}\right)\bar{n}^\mu,
\end{eqnarray}
here $E$ is the off-shell energy \cite{Ebert:2001pc} of the order $(m_B/2)$ while the momenta satisfy the on-shell conditions, i.e., $p_B^2=m_B^2$, $p_T^2=m_T^2$ and $p_\gamma^2=0$, respectively.  This will help us to define the momenta of quarks and gluons according to different momentum regions in the heavy-quark limit, e.g., the soft region scales like $p_B-m_bv\sim\Lambda_{\text{QCD}}$.  Generally, we decompose the momentum in terms of the light cone variables as
\begin{equation}
p^\mu=n\cdot p \frac{\bar{n}^\mu }{2}+\bar{n}\cdot p \frac{n^\mu}{2}+p_\perp^\mu=p_+^\mu +p_-^\mu +p_\perp^\mu ,
\end{equation}
where the different regions will thus have varying components of $p_+^\mu$, $p_-^\mu$ and $p_\perp^\mu$. Table \ref{TABLE1} summarizes the regions and their scaling for different momentum labels, and we require them to satisfy the factorization relation given in  Eq. (\ref{theorem}) for these heavy to light radiative decays. The collinear region is further divided into $n$ and $\bar{n}-$collinear regions, although the later do not appear for the radiative decays because $q^2\equiv (p_B-p_T)^2=p_\gamma^2=0$. Two more regions, i.e., hard and hard-collinear (again divided into $n$ and $\bar{n}-$ hard-collinear) are required to correctly reproduce the QCD infrared physics, manifested in the WCs $\mathcal{C}^{I}$ and jet functions, respectively. Another possible mode is the soft-collinear mode, and this is interesting because the interaction of soft with a soft-collinear will yield a soft mode, while the interaction of collinear with soft-collinear will yield a collinear mode. However, the interaction of a soft with a collinear mode will yield a hard-collinear mode making the factorization relation unreliable. Therefore, to have the factorization valid, one has to show that up to any order in $\lambda$ these modes do not contribute. Taking outgoing collinear quarks to be massless is also valid assumption for our case. Now, if the collinear quarks have a mass of order $\Lambda_{\text{QCD}}$, then the soft-collinear mode disappear - but in this case the soft and collinear propagators are not well-defined and require additional regulators. This may further require to prove factorization that is  independent of these regulator, but it is not required in our case because $m_{s,d}<\Lambda_{\text{QCD}}$.
\begin{table}[t]
 \caption{Different momentum regions and their scaling  that may appear in heavy-quark limit using the expansion parameter $\lambda\sim\Lambda_{\text{QCD}}/E$.}\label{TABLE1}
 \centering
    \begin{tabular}{|l|l|l|}
    \hline
    \textbf{Label}&\quad\textbf{Region}&\ \textbf{Scaling} \\  \hline
     Hard&\quad(1,1,1)&\quad  $m_b$ \\  \hline
      Hard Collinear&\quad$(\lambda,1,\lambda^{1/2})$& \quad  $m_b\lambda^{1/2}$ \\  \hline
      Collinear&\quad$(\lambda^2,1,\lambda)$&\quad  $m_b\lambda$ \\  \hline
      Soft&\quad$(\lambda,\lambda,\lambda)$&\quad $m_b\lambda$ \\  \hline
      Soft-Collinear&\quad$(\lambda^2,\lambda,\lambda^{3/2})$\ \ &\quad $m_b\lambda^{3/2}$ \\  \hline
     \end{tabular}
\end{table}

\section{QCD to SCET$_{\text{I}}$ matching}\label{sec2}

The hard-collinear quark and gluon fields along with heavy-quark and soft fields scale as
\begin{eqnarray}
\xi_{hc} , \xi_{\bar{hc}} \sim\lambda^{1/2},\quad A_{hc}^\mu \sim (\lambda , 1, \lambda^{1/2}),\quad h\sim \lambda^{3/2},\quad A_s\sim(\lambda,\lambda,\lambda).\label{Q}
\end{eqnarray}
In Eq. (\ref{Q}), $\xi,\; A^\mu$ and $h$ denote the quark, gluon and the heavy-quark fields, respectively. The subscripts $hc,\; s$ and $c$ represent the hard-collinear, soft and collinear modes, respectively. In SCET, the operators with derivatives of hard-collinear or collinear field are not suppressed for the larger momentum component and they can have same power counting. Therefore, to have gauge-invariant operators, the fields are connected by light-like Wilson's lines at different points, e.g., in SCET$_\text{I}$, the hard-collinear fields have to be invariant under hard-collinear gauge transformations, and this requires to introduce a hard-collinear Wilson line
\begin{align}
\mathcal{W}_{hc}(x)=\mathbb{P}\ \text{exp}\left(ig\int_{-\infty} ^0 ds_1\  \bar{n}\cdot A_{hc}(x+s_1\bar{n})\right), \label{Wiline}
\end{align}
where the gauge-invariant hard-collinear fields are defined as
\begin{align}
\chi_{hc}(x)&=W^\dagger_{hc}(x)\xi_{hc}(x)\nonumber\\
\mathcal{A}_{hc} ^\mu (x)&=W^\dagger _{hc}[iD_{hc}^\mu (x)W_{hc}(x)]+\frac{\bar{n}^\mu}{2}[W^\dagger _{hc}(x)gn\cdot A_s(x_-)W_{hc}(x)-gn\cdot A_s(x_-)].
\end{align}
In above equations, it can be noticed that the position variable, $x$, is defined as $x\equiv x_+ +x_- +x_\perp$ and $s_1$ is the light-ray variable in Eq. (\ref{Wiline}).  The mass dimensions are used to construct the operators in SCET$_{\text{I}}$ \cite{Becher:2005fg}. The simplest is the two-particle operator $(J^{\mathscr{A}})$ at dimension-3 while the higher power (subleading) three-particle operator $(J^{\mathscr{B}})$ has dimension-4. The $\mathscr{A}$-type operators are invariant under first re-parameterization condition that takes into account invariance under rescaling of the light-like vectors, i.e., $n^\mu \rightarrow (1+\alpha)n^\mu$ and $\bar{n}^\mu \rightarrow (1-\alpha)\bar{n}^\mu$. The operators are constructed using scalar $\left(\Gamma= I\right)$, vector $\left(\Gamma_i = (\gamma ^\mu, v^\mu , n^\mu )\right)$, and tensor $\left(\Gamma_j= \gamma^{[\mu}\gamma^{\nu]},\gamma^{[\mu}v^{\nu]},\gamma^{[\mu}n^{\nu]},v^{[\mu}n^{\nu]}\right)$ Dirac structures, where square brackets denote the anti-commutation\cite{Becher:2005fg,Hill:2004if,Beneke:2004rc}, and subscripts $i=1,\cdots,3$ and $j=1,\cdots, 4$, correspond to three vector and four-tensor structures, respectively.  Similarly, we can construct the $\mathscr{B}$-type operators, but since these are subleading, the Dirac structures required to be invariant under the second reparameterization condition allowing the light-like vectors to remain invariant in the perpendicular components, i.e., $n^\mu \rightarrow n^\mu +\epsilon_\perp^\mu$ and $\bar{n}^\mu\rightarrow \bar{n}^\mu -\epsilon_\perp^\mu$. This will mix the $\mathscr{B}$-type operators and one needs to define a new basis (for details see ref. \cite{Hill:2004if}). In the change of basis, we now redefine the gamma matrices as $\gamma_\perp ^\mu = \gamma^\mu-n^\mu\slashed{\bar{n}}/2-\bar{n}^\mu\slashed{n}/2$.  Thus, the new $\mathscr{B'}$-type operators could have the Dirac structure $\Gamma_{i}^{'\mu}=(\gamma_\perp^\mu,v^\mu ,n^\mu)$ for vectors and $\Gamma_j ^{'\mu\nu}=(\gamma^{[\mu}_\perp \gamma^{\nu]}_\perp,\gamma^{[\mu}_\perp v^{\nu]},\gamma^{[\mu}_\perp n^{\nu]},v^{[\mu}n^{\nu]})$ for tensors. The tree-level WCs for the two types $\left(\mathscr{A},\; \mathscr{B}\right)$ of operators introduced above are
\begin{align}
J_{S}^{\mathscr{A}}&=\bar{\chi}_{hc}(s_1\bar{n})h(0),\qquad\qquad\qquad\qquad\qquad\qquad\qquad C_S^{\mathscr{A}}=1,\nonumber\\
J_{V_i}^{\mathscr{A}}&= \bar{\chi}_{hc}(s_1\bar{n})\Gamma_{i}h(0),\quad\qquad\qquad\qquad\qquad\qquad\qquad C_{V1}^{\mathscr{A}}=1,C_{V2}^{\mathscr{A}}=0,C_{V3}^{\mathscr{A}}=0,\nonumber\\
J_{S}^{\mathscr{B}'}&=\bar{\chi}_{hc}(s_1\bar{n})\slashed{\mathcal{A}}_\perp(r\bar{n})h(0),\ \ \qquad\qquad\qquad\qquad \ \  \  \ \ \  C_S ^{\mathscr{B}'}=-1,\nonumber\\
J_{Vi}^{\mathscr{B}'\mu}&=\bar{\chi}_{hc}(s_1\bar{n})\slashed{\mathcal{A}}_\perp(r\bar{n})\Gamma_i^\mu h(0),\ \ \  \ \qquad\qquad\qquad\qquad C_{V1}^{\mathscr{B}'}=1,C_{V2}^{\mathscr{B}'}=-2,C_{V3}^{\mathscr{B'}}=1-z,\nonumber\\
J_{V4}^{\mathscr{B}'\mu}&=\bar{\chi}_{hc}(s_1\bar{n})\gamma_\perp ^\mu\slashed{\mathcal{A}}_\perp(r\bar{n}) h(0),\ \ \  \ \qquad\qquad\qquad\qquad C_{V4}^{\mathscr{B}'}=0,\nonumber\\
J_{Tj}^{\mathscr{B}'\mu}&=\bar{\chi}_{hc}(s_1\bar{n})\slashed{\mathcal{A}}_\perp(r\bar{n})\Gamma_j^{\mu\nu} h(0),\ \qquad\qquad\qquad\qquad\  \ C_{T1}^{\mathscr{B}'}=-1, C_{T2}^{\mathscr{B}'}=-4,C_{T3}^{\mathscr{B}'}=2,C_{T4}^{\mathscr{B}'}=2,\nonumber\\
J_{T5}^{\mathscr{B}'\mu}&=\bar{\chi}_{hc}(s_1\bar{n})\mathcal{A}_{\perp\alpha}(r\bar{n})\gamma_\perp^{[\alpha}\gamma^\mu_\perp\gamma^{\nu]}_\perp h(0),\qquad \qquad\qquad  \ C_{T5}^{\mathscr{B}'}=0\nonumber\\
J_{T6}^{\mathscr{B}'\mu}&=\bar{\chi}_{hc}(s_1\bar{n})v^{[\mu}\gamma_\perp^{\nu]}\slashed{\mathcal{A}}_{\perp}(r\bar{n}) h(0)\qquad\qquad\qquad\qquad  \ C_{T6}^{\mathscr{B}'}=0,\nonumber\\
J_{T7}^{\mathscr{B}'\mu}&=\bar{\chi}_{hc}(s_1\bar{n})n^{[\mu}\gamma_\perp^{\nu]}\slashed{\mathcal{A}}_{\perp}(r\bar{n}) h(0),\ \quad\qquad\qquad\qquad  \  C_{T7}^{\mathscr{B}'}=2z.\label{tlevelC}
\end{align}
The WCs for the corresponding pseudo-scalar and pseudo-tensor operators will remain the same, while those of the axial-vector currents will be
\begin{align}
J_{A_i}^{\mathscr{A}\mu}&= \bar{\chi}_{hc}(s_1\bar{n})\Gamma_{i}^\mu\gamma_5 h(0),\qquad\qquad\qquad\qquad\quad\ C_{A1}^{\mathscr{A}}=1,C_{A2}^{\mathscr{A}}=0,C_{A3}^{\mathscr{A}}=0,\nonumber\\
J_{Ai}^{\mathscr{B}'\mu}&=\bar{\chi}_{hc}(s_1\bar{n})\slashed{\mathcal{A}}_\perp(r\bar{n})\Gamma_i^\mu \gamma_5h(0),\qquad\qquad\qquad C_{A1}^{\mathscr{B}'}=-1,C_{A2}^{\mathscr{B}'}=-2,C_{A3}^{\mathscr{B'}}=1-z,\nonumber\\
J_{A4}^{\mathscr{B}'\mu}&=\bar{\chi}_{hc}(s_1\bar{n})\gamma_\perp ^\mu\gamma_5\slashed{\mathcal{A}}_\perp(r\bar{n}) h(0),\quad\qquad\qquad\ \ \  C_{A4}^{\mathscr{B}'}=0,\label{BtypetlevelC}
\end{align}
where $z=2E/m_b$ with $E$ being the energy of the final state quark. According to the factorization introduced in Eq. (\ref{theorem}), the matching at leading power from QCD to SCET$_{\text{I}}$ can be written as
\begin{align*}
\mathcal{H}_{\text{eff}}&\rightarrow -\left(\int ds \int da\ \tilde{C}_i^{\mathscr{A}}(s)J_i ^{\mathscr{A}}(s)\right.\nonumber\\
&\quad +\sum_{j=1,2}\left.\int ds \int dr \int da\ \tilde{C}^{\mathscr{B'}}_j (s,r,a)J_j^{\mathscr{B'}}(s,r,a)\right)+\cdots.
\end{align*}
For different Dirac structures, the relevant operators for $b\rightarrow (s,d)\gamma$ decay at leading power in $\lambda$ along with their WCs can be obtained from Eqs.(\ref{tlevelC}) and (\ref{BtypetlevelC}).
%To construct the operators in SCET$_{\text{I}}$ for $B\rightarrow A\gamma$, it is required to have one $n$-hard-collinear field, one $\bar{n}$-hard-collinear field ($\gamma$ in our case) and the heavy quark field. So at leading power in $\lambda$ the relevant operator and WC for A-type operator is  After introducing the 2-point operators ($\mathscr{A}$-type) and 3-point operators ($\mathscr{B}$-type), we need to consturct the leading power SCET$_{\text{I}}$ operators corresponding to QCD $b\rightarrow (s,d)\gamma$. 
The $\mathscr{A}-$type  operator and its WC is
\begin{eqnarray}
J^{\mathscr{A}} (s_1,a) &=& \bar{\chi}_{hc}(s_1\bar{n})(1+\gamma_5)\slashed{\mathcal{A}}_{\bar{hc}\perp}^{(em)} h(0)\nonumber\\ \label{Aoper}
C^{\mathscr{A}}(E,E_\gamma)&=&\int ds_1\int da e^{is\bar{n}\cdot p}e^{ian\cdot p_\gamma}\tilde{C}^{\mathscr{A}}(s,a),
\end{eqnarray}
 where $E =  \bar{n}\cdot p/2$ and $E_\gamma\equiv n\cdot p_\gamma /2$. Also, $\bar{n}\cdot p$ is the large component of the total outgoing $n$-hard-collinear momentum and $n\cdot p_\gamma$ is the momentum of an outgoing photon. Similarly, the relevant SCET$_{\text{I}}$ operators for $\mathscr{B'-}$type currents and their corresponding WCs are
\begin{eqnarray}
J^{\mathscr{B'}}_1(s_1,r,a)&=&\bar{\chi}_{hc}(s_1\bar{n})(1+\gamma_5)\slashed{\mathcal{A}}_{\bar{hc}\perp}^{(em)}(an)\slashed{\mathcal{A}}_{hc\perp}(r\bar{n}) h(0)\label{Boper1}\\
J^{\mathscr{B'}}_2(s_1,r,a)&=&\bar{\chi}_{hc}(s_1\bar{n})(1+\gamma_5)\slashed{\mathcal{A}}_{hc\perp}(r\bar{n})\slashed{\mathcal{A}}_{\bar{hc}\perp}^{(em)}(an) h(0)\label{Boper2}\\
C^{\mathscr{B'}} _j(E,E_\gamma,u)&=&\int ds_1\int dr\int da e^{i(us_1 +\bar{u}r)\bar{n}\cdot p}e^{ian\cdot p_\gamma}\tilde{C}_j ^{\mathscr{B'}}(s_1,r,a).
\end{eqnarray}
The index $j$ in $\mathscr{B'-}$type WCs refer to the two types of operators represented by Eqs. (\ref{Boper1}) and Eq. (\ref{Boper2}). The variable $u(\bar{u})$ denotes the fraction of the large component of the $n$-hard-collinear momentum carried by the quark (gluon) field.

There could be a contribution from another set of diagrams with photon-emission from the spectator quark. The relevant SCET$_{\text{I}}$ operator for these diagrams ($\mathscr{C}-$type matching) is a four-quark operator. Together with its WC, we can write
\begin{eqnarray}
J^{\mathscr{C}}(s_1,r,a)&=&\bar{\chi}_{hc}(s_1\bar{n})(1+\gamma_5)\frac{\slashed{\bar{n}}}{2}\chi_{hc}(r\bar{n})\bar{\chi}_{\bar{hc}}(an)(1+\gamma_5)\frac{\slashed{n}}{2} h(0)\nonumber\\\label{Coper1}
C_k ^{\mathscr{C}}(u)&=&\int ds_1\int dr\int da e^{i(us_1 +\bar{u}r)\bar{n}\cdot p}e^{ian\cdot p_\gamma}\tilde{C}_k ^{\mathscr{C}}(s_1,r,a).
\end{eqnarray}

\begin{figure*}[t!]
    \centering
    \begin{subfigure}[t]{0.3\textwidth}
        \centering
        \includegraphics[height=1in]{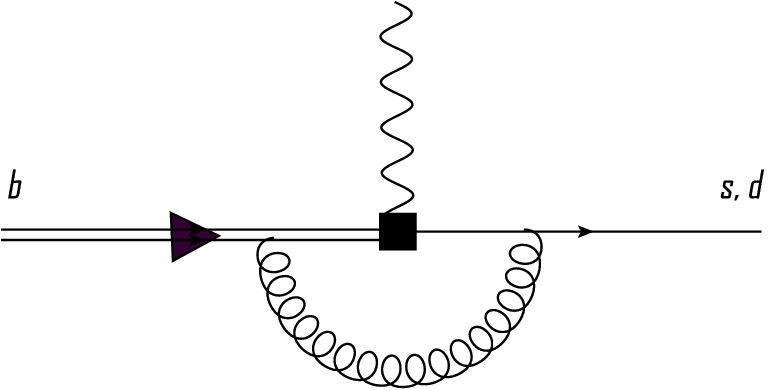}
        \caption{}
        \label{q7a}
    \end{subfigure}%
    ~ 
    \begin{subfigure}[t]{0.3\textwidth}
        \centering
        \includegraphics[height=1in]{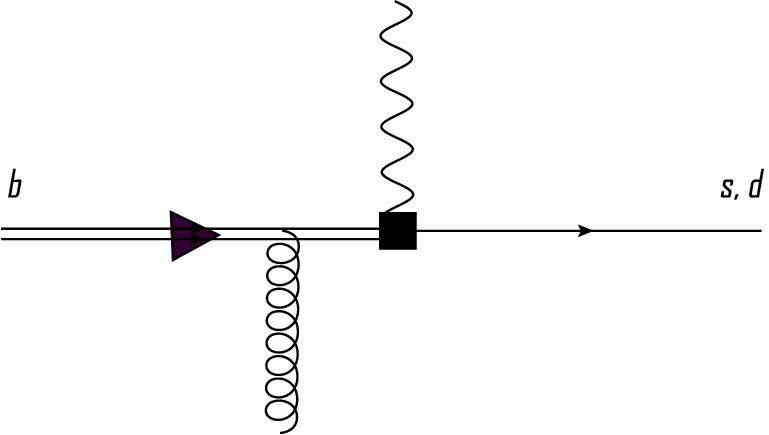}
        \caption{}
        \label{q7b1}
    \end{subfigure}
    ~ 
    \begin{subfigure}[t]{0.3\textwidth}
        \centering
        \includegraphics[height=1in]{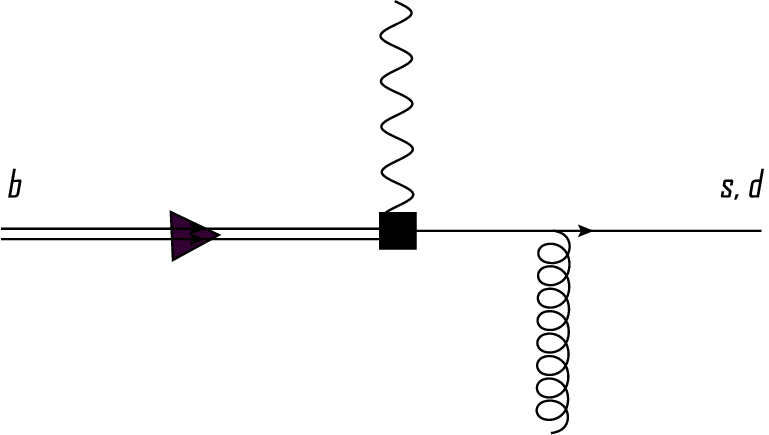}
        \caption{}
        \label{q7b2}
    \end{subfigure}
    \caption{QCD diagrams for $Q_7$ operator matching onto $J^{\mathscr{A}}$,$J_1^{\mathscr{B'}}$ and $J_2^{\mathscr{B'}}$, respectively \cite{Sikandar:2021shv}.}
\end{figure*}

The matching calculation at an order of $\alpha_s$ for $\mathscr{A}-$type operator is a loop diagram while for $\mathscr{B'}-$type it is at a tree level. 
Let us denote $\Delta_i C_j^{(\mathscr{A},\mathscr{B}^\prime,\mathscr{C})}$ as the matching result of weak effective operators $Q_i$ to the SCET$_{\text{I}}$ operators $J_j ^{\mathscr{A},\mathscr{B'},\mathscr{C}}$, and write the total matching coefficients as
\begin{eqnarray}
\mathcal{C}^{\mathscr{A}}&=&\frac{G_F}{\sqrt{2}}\sum_{p=u,c}V_{ps}^{*}V_{pb}\left[C_1(\mu_Q )\Delta_1 ^p C^{\mathscr{A}}(\mu_Q ,\mu) + C_7(\mu_Q )\Delta_7  C^{\mathscr{A}}(\mu_Q ,\mu)+C_8(\mu_Q )\Delta_8  C^{\mathscr{A}}(\mu_Q ,\mu) \right],\nonumber\\
\mathcal{C} ^{\mathscr{B'}}&=&\frac{G_F}{\sqrt{2}}\sum_{p=u,c}V_{ps}^{*}V_{pb}\left[C_1(\mu_Q )\Delta_1 ^p C^{\mathscr{B'}}_{1,2}(\mu_Q ,\mu) +C_7(\mu_Q )\Delta_7  C^{\mathscr{B'}}_{1,2}(\mu ,\mu)+C_8(\mu_Q )\Delta_8  C^{\mathscr{B'}}_{1,2}(\mu_Q ,\mu)  \right],\nonumber\\
\mathcal{C}^{\mathscr{C}}&=&\frac{G_F}{\sqrt{2}}\sum_{p=u,c}V_{ps}^{*}V_{pb}\left[C_1(\mu_Q )\Delta_1 ^p C^{\mathscr{C}}(\mu_Q ,\mu) + C_7(\mu_Q )\Delta_7  C^{\mathscr{C}}(\mu_Q ,\mu)+C_8(\mu_Q )\Delta_8  C^{\mathscr{C}}(\mu_Q ,\mu) \right].\nonumber\\\label{WCmatching}
\end{eqnarray}
Here $\mu_Q $ is the scale at which QCD is matched to SCET$_{\text{I}}$ and $\mu$ is the renormalization scale in the effective theory. The matching of $Q_7$ is performed at loop-level (Fig\ref{q7a} ) for $\mathscr{A}-$type ($J^{\mathscr{A}}$) and at tree-level (Fig(\ref{q7b1},\ref{q7b2})) for $\mathscr{B}-$type ($J^{\mathscr{B}^\prime}$) operators \cite{Beneke:2004rc,Bauer:2000yr,Beneke:2015wfa}:  %The tree level matching of $Q_7$ is trivial as can be seen from Eq. (\ref{treelev}). The matching coefficient at order $\mathcal{O}(\alpha_s)$ can be obtained via matching the QCD diagram \ref{q7a} to SCET$_{\text{I}}$ current $J^A$ as given in \cite{Beneke:2004rc,Bauer:2000yr}

\begin{eqnarray}
\Delta_7 C^\mathscr{A} &=&\frac{e\bar{m}_b E_\gamma}{2\pi^2}\left\lbrace 1 +\frac{\alpha_s(m_b)C_F}{4\pi}\left[-2 \text{ln}^2\frac{\mu }{2E} -5\text{ln}\frac{\mu}{2E}-2\text{ln}\frac{\mu_{_{QCD}}}{2E}- 2\text{Li}_2(1-\frac{2E}{m_b})-6-\frac{\pi^2}{12}\right]\right\rbrace,\nonumber\\
\Delta_7 C_1^\mathscr{B'}&=&\frac{e\bar{m}_b E_\gamma}{4\pi^2 m_b},\nonumber\\
\Delta_7 C_2^\mathscr{B'}&=&0.
\end{eqnarray}
Here $\bar{m}_b$ is the $b-$quark mass at next-to-leading order, i.e., 
\begin{eqnarray}
\bar{m}_b&=& m_b\left[1+\frac{\alpha_s(\mu) C_F}{4\pi}\left(3\text{ln}\frac{m_b^2}{\mu^2}-4\right)\right].\nonumber
\end{eqnarray}

\begin{figure*}[t!]
    \centering
    \begin{subfigure}[t]{0.3\textwidth}
        \centering
        \includegraphics[height=1in]{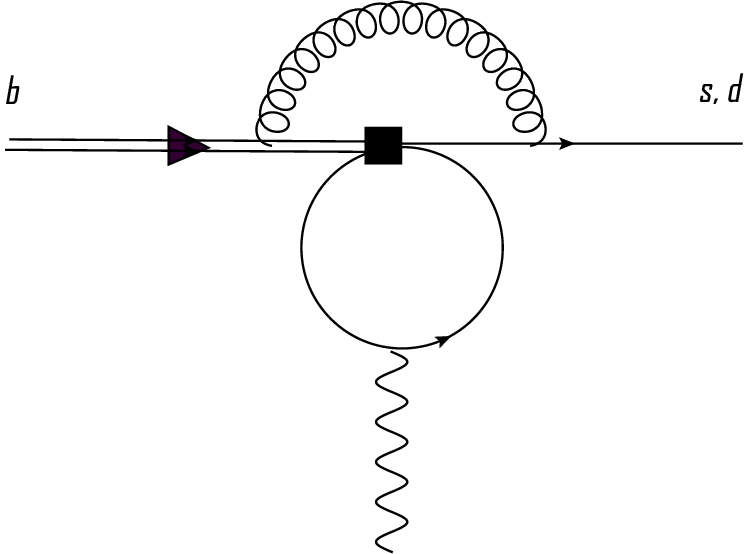}
        \caption{}
        \label{q1a}
    \end{subfigure}%
    ~ 
    \begin{subfigure}[t]{0.3\textwidth}
        \centering
        \includegraphics[height=1in]{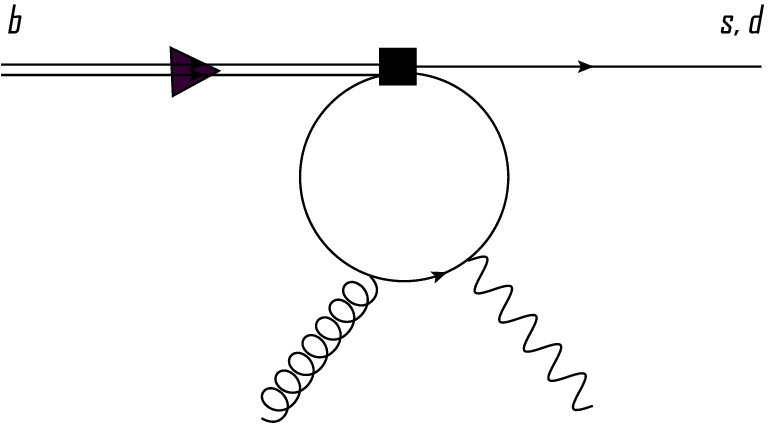}
        \caption{}
        \label{q1b1}
    \end{subfigure}
    ~ 
    \begin{subfigure}[t]{0.3\textwidth}
        \centering
        \includegraphics[height=1in]{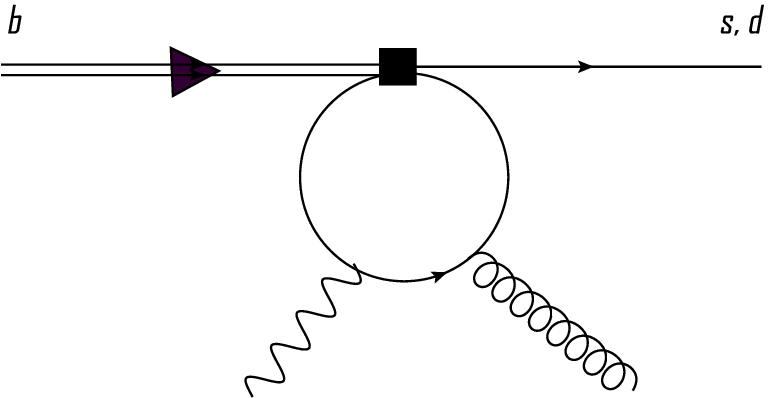}
        \caption{}
        \label{q1b2}
    \end{subfigure}
    \caption{QCD diagrams for $Q_1$ operator matching onto $J^{\mathscr{A}}$,$J_1^{\mathscr{B'}}$ and $J_2^{\mathscr{B'}}$, respectively \cite{Sikandar:2021shv} .}
\end{figure*}

The matching of $Q_1$ operator with a charm quark loop (setting $m_u=0$) for $\mathscr{A}-$type matching (Fig (\ref{q1a})), and $\mathscr{B'}-$type matching (Fig (\ref{q1b1},\ref{q1b2})) give the following WCs
%For the operator $Q_1$ with a $c-$quark loop (setting $m_u=0$), the QCD diagram that contributes at leading order for the $\mathscr{A}-$type current matching is given in Fig. \ref{q1a}. It is a vertex diagram from which hard modes are to be integrated out along with the contribution of the fermion loop when matching to SCET$_{\text{I}}$. Figs. \ref{q1b1} and \ref{q1b2} are bremsstrauhlung diagrams with an on-shell photon. The corresponding coefficients are
\begin{eqnarray}
\Delta_1 C^\mathscr{A} &=&\frac{\alpha_s C_F}{4\pi}G_i(x_c)\Delta_7 C^\mathscr{A}, \nonumber\\
\Delta_1 C^\mathscr{B'} _1 (u) &=&-\Delta_1 ^q C_2 ^\mathscr{B'}(u)=\frac{2e}{3}f\left(\frac{\bar{m}_c ^2}{4\bar{u}EE_\gamma}\right) \Delta_7 C_1 ^\mathscr{B'},
\end{eqnarray}
where $x_c =\bar{m}_c ^2/m_b^2$ and the functions $G_1(x_c)$ and $f\left(\frac{m_c^2}{4EE_\gamma\bar{u}}\right)$ are given in Appendix.

\begin{figure*}[t!]
    \centering
    \begin{subfigure}[t]{0.3\textwidth}
        \centering
        \includegraphics[height=1in]{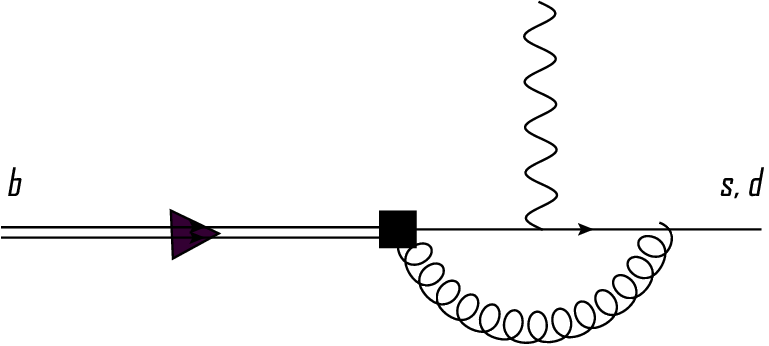}
        \caption{}
        \label{q8a}
    \end{subfigure}%
    ~ \qquad
    \begin{subfigure}[t]{0.3\textwidth}
        \centering
        \includegraphics[height=1in]{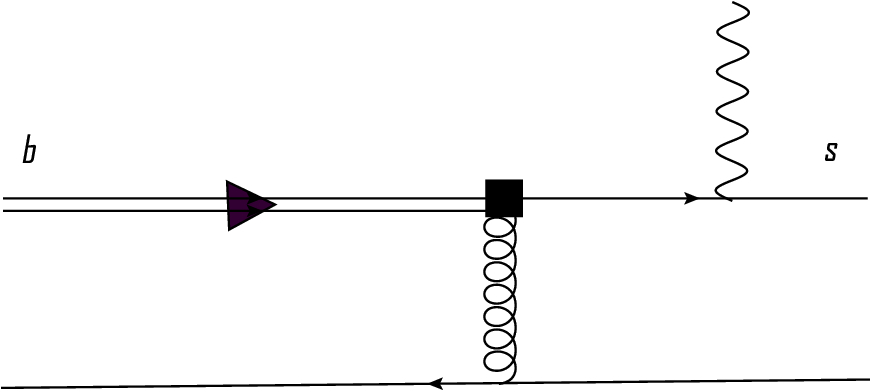}
        \caption{}
        \label{q8b}
    \end{subfigure}
    \caption{QCD diagrams for $Q_8$ operator matching onto $J^{\mathscr{A}}$,$J_1^{\mathscr{B'}}$ and $J_2^{\mathscr{B'}}$, respectively \cite{Sikandar:2021shv}.}
\end{figure*}

The relevant diagrams for matching the chromomagnetic operator $Q_8$ to SCET$_{\text{I}}$ $\mathscr{A}-$ and $\mathscr{B'}-$types operators are given in Figs. \ref{q8a} and \ref{q8b}, respectively. The diagrams where a photon is emitted from the $b-$quark for $Q_8$ operator are suppressed. The WCs thus found are \cite{Greub:1996tg}:% Diagrams in contrast to Fig. \ref{fig:fig3}, where the photon is emitted from the $b-$quark are suppressed, so $\Delta_8 C_2 ^B (u)\simeq0$. The coefficients for the A- and B- type currents matching are \cite{Greub:1996tg}
\begin{eqnarray}
\Delta_8 C^\mathscr{A}&=&\frac{\alpha_s C_F}{4\pi}G_8 \Delta_7 C^\mathscr{A},\nonumber\\
\Delta_8 C_1 ^\mathscr{B'}(u)&=& \frac{\bar{m}_b}{4\pi^2}\frac{e}{3}\frac{\bar{u}}{u},\nonumber\\
\Delta_8 C_2 ^\mathscr{B'} (u)&=& 0.
\end{eqnarray}

There are a few other diagrams in which the photon can be emitted but they are power suppressed. One such diagram is the photon emission from $b-$quark for $Q_1$ matching onto $J^{\mathscr{A}}$. Another is the photon emission from the outgoing spectator quark which is also power-suppressed \cite{Chay:2003kb,DescotesGenon:2004hd}. Although the photon emission from incoming soft-spectator quark are eventually not power suppressed when matched to SCET$_{\text{II}}$, but they vanish when the matrix elements are calculated for the transversely polarized tensor meson.

\section{SCET$_{\text{II}}$ Operators and SCET$_{\text{I}}$ $\rightarrow$ SCET$_{\text{II}}$ matching}\label{sec4}
In the two step matching procedure, the next step is to match the intermediate theory SCET$_\text{I}$ to the final theory SCET$_\text{II}$ to find the corresponding WCs, which are also known as the jet functions. To start with, let us define the gauge-invariant fields required for SCET$_\text{II}$, i.e., 
\begin{eqnarray}
\bar{\chi}_c &=& \mathcal{W}_c ^{\dagger}\xi_c \sim \lambda ,\qquad Q_s = \mathcal{S}^\dagger q_s \sim \lambda ^{3/2}, \qquad \mathcal{H}= \mathcal{S}^\dagger h \sim \lambda^{3/2} \nonumber\\
\mathcal{A}_c ^\mu &=& \mathcal{W}_c ^{\dagger}(i D_c ^\mu \mathcal{W}_c)\sim (\lambda^2 ,0,\lambda), \quad 
 \mathcal{A}_s ^\mu = \mathcal{S} ^{\dagger}(i D_s ^\mu \mathcal{S})\sim (\lambda^2 ,0,\lambda).
\end{eqnarray}
Here the gauge-invariant operators are constructed using the soft and collinear Wilson lines 
%To define the gauge-invariant operators, once again we need to define soft and collinear Wilson lines as they were defined for hard-collinear fields in Eq. (\ref{Wiline});
\begin{eqnarray}
\mathcal{W}_{c}(x)&=\mathbb{P}\ \text{exp}\left(ig\int_{-\infty} ^0 ds_1 \bar{n}\cdot A_{c}(x+s_1\bar{n})\right)\nonumber\\
\mathcal{S}(x)&= \mathbb{P}\ \text{exp}\left(ig\int_{-\infty} ^0 ds_1 \bar{n}\cdot A_{s}(x+s_1\bar{n})\right).
\end{eqnarray}
 The non-factorizable part (soft-overlap function) can be defined as the matrix elements of $J^\mathscr{A}$ operators, therefore, the $\mathscr{A}-$type matching of SCET$_\text{I}$ to SCET$_\text{II}$ is not required. For $\mathscr{B'}-$type matching, the SCET$_\text{II}$ operators at tree level can be defined as
%We are not required to perform SCET$_{\text{I}}$ to SCET$_{\text{II}}$ matching for A-type currents as the matrix elements of $J^A$ already give the non-factorizable part i.e., soft-overlap function.  Therefore, we only need to define B- types operators for matching. Recall, there were two B-type currents, so they match onto as
\begin{eqnarray}
O_1 ^{\mathscr{B'}}(s_1,t)&=&\bar{\chi}_c (s_1\bar{n})(1+\gamma_5)\slashed{\mathcal{A}}_{\bar{c}\perp}^{(em)}(0)\frac{\bar{\slashed{n}}}{2}\chi_c (0)\bar{\mathcal{Q}}_s (tn)(1-\gamma_5)\frac{\slashed{n}}{2}\mathcal{H}_s (0)\nonumber\\
O_2 ^{\mathscr{B'}}(s_1,t)&=&\bar{\chi}_c (s_1\bar{n})(1+\gamma_5)\frac{\bar{\slashed{n}}}{2}\chi_c (0)\bar{\mathcal{Q}}_s (tn)(1+\gamma_5)\frac{\slashed{n}}{2}\slashed{\mathcal{A}}_{\bar{c}\perp}^{(em)}(0)\mathcal{H}_s (0).\label{scet2op}
\end{eqnarray}

\begin{figure}
\centering
  \includegraphics[width=100mm]{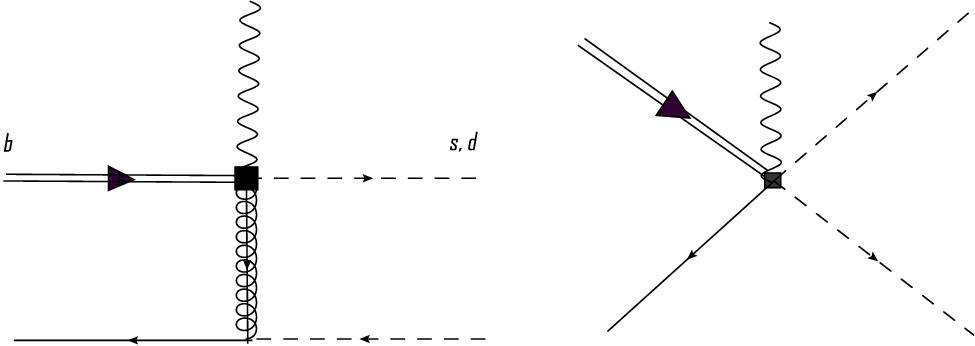}
\caption{SCET$_{\text{I}}$ diagram matching onto SCET$_{\text{II}}$ 4-quark operator. The dashed gluon is hard-collinear\cite{Sikandar:2021shv}.}
\label{fig:fig6}
\end{figure}

%The photon field is in the $\bar{n}$ direction and is collinear. The first operator matches SCET$_{\text{I}}$ operator with photon emission from $b-$quark and it requires opposite chirality of the spectator quark field, $\bar{\mathcal{Q}}_s$. On the other hand, the second operator matches SCET$_{\text{I}}$ operator with photon emission from $s-$quark.
From Eq. (\ref{scet2op}), one can notice that the leading power operators have a collinear and soft part. Thus in SCET$_\text{II}$, the matrix elements for $B\rightarrow T$ can be written in terms of the LCDAs, independently for soft $B$ and collinear tensor mesons. This decoupling into soft and collinear parts prevented end-point divergences to  appear in the convolution integrals of Eq. (\ref{theorem}).
%It can be noticed that at leading power, the SCET$_{\text{II}}$ operators have a soft and a collinear part only. Therefore, the matrix elements will be independently found in terms of LCDA of $B-$ and final state axial-vector mesons. The decoupling of soft-collinear modes made our factorization successful. Moreover, if they were not decoupled, they would have appeared as end-point divergences in convolution integrals (\ref{theorem}) which is in contrast to the LEET as discussed in \cite{Sikandar:2019qyb}. Just to mention, the LEET does not correctly produce the infrared divergences, therefore, we were left with end-point divergences arising from the convolution integral at leading twist. These end-point divergences were then assumed to be absorbed in soft-form factor in the LEET. Hence, it would suggest the difference of soft-form factors in both theories.
The corresponding WC in momentum space is given as
\begin{equation}
D_{1,2} ^{\mathscr{B'}}(\omega, u)\equiv \int ds_1 \int dt e^{-i\omega n\cdot v t}e^{ius_1\bar{n}\cdot p}\tilde{D}_{1,2} ^\mathscr{B'} (s_1,t),
\end{equation}
which is a convolution of the SCET$_{\text{I}}$ WC $C_i ^B$, with a jet function $\mathcal{J}_{\perp,\parallel}$, i.e.,
\begin{eqnarray}
D_{1,2} ^\mathscr{B'}(\omega, u, \mu_i)=\frac{1}{\omega}\int_0 ^1 dy \mathcal{J}_{\perp} \left(u,y,\text{ln}\frac{2E\omega}{\mu_i ^2 },\mu_i\right)C_{1,2} ^\mathscr{B'}(y,\mu_1),
\end{eqnarray}
where $\mu_i\sim \sqrt{2E\Lambda_{QCD}}$ is an intermediate scale. At tree level, the matching of $J_1 ^\mathscr{B'}$ onto $O_1 ^\mathscr{B'}$ is trivial (c.f. Fig. \ref{fig:fig6}) and the corresponding jet function is given as
\begin{eqnarray}
\mathcal{J}_\perp (u,v)&=&\mathcal{J}_\parallel (u,v)= -\frac{4\pi C_F \alpha_s}{N}\frac{1}{2E\bar{u}}\delta(u-v)
\end{eqnarray}
while the similar matching for $J_2 ^\mathscr{B'}$ is suppressed at a leading power. 
\section{Resummation}\label{sec5}
In the matching calculations of previous sections, we introduced two scales, i.e., the hard scale $\mu_Q\sim m_b$ at which we matched QCD to SCET$_\text{I}$ and intermediate scale $\mu_i\sim \sqrt{2E\Lambda_{QCD}}$ at which SCET$_\text{I}$ is matched to SCET$_\text{II}$. To evolve the matching coefficients from a higher to a lower scale, we are required to solve the renormalization group (RG) equation for SCET$_\text{I}$ operators. In this context, since only the hard scale $\mu_Q$ is involved in the matching calculations for the $\mathscr{A-}$type operators, we may choose the dependence of soft overlap function on $\mu_Q$, which help us to avoid the RG evolution of $\mathscr{A-}$type operators. For the $\mathscr{B'-}$type operators, the evolution of scale requires the calculation of the one-loop anomalous dimension by keeping the UV divergent terms appearing in SCET$_\text{I}$ loop diagrams (c.f. Fig. 3 of \cite{Hill:2004if}).

%The WCs calculated in previous section are scale dependent, and hence their validity expected to a hard-scale $\approx m_b$. Therefore, one needs to re-sum the large logarithms in corrected coefficients at an intermediate scale. Principally , one needs to separately re-sum the logarithms in the matching calculation of QCD to SCET$_{\text{I}}$ and then from SCET$_{\text{I}}$ to SCET$_{\text{II}}$. In this context, the $\mathscr{A-}$type coefficients $\mathcal{C}^\mathscr{A}$ are obtained at hard scale and their contribution is quite small to neglect. The running of $\mathscr{B-}$ type operators requires the calculation of the one-loop anomalous dimension by keeping the UV divergent terms appearing in SCET$_{\text{I}}$ loop diagrams (c.f. Fig. 3 of \cite{Hill:2004if}). %As we are dealing with tensor mesons whose mass is around $1.3$ GeV, and lies very close to the intermediate scale i.e., around $1.5$ GeV, therefore, the resummation of SCET$_{\text{II}}$ operators is not needed in our case.%%%
The evolution of the $\mathscr{B'}-$type coefficients read as \cite{Sikandar:2022iqc}
\begin{eqnarray}
\frac{d}{d\text{ln}\mu}\mathcal{C}^\mathscr{B'} _j (E,v)&=&\gamma^\mathscr{B'} _{ij}(u,v)\mathcal{C} ^\mathscr{B'} _i(E,u),\label{anomalyb}
\end{eqnarray}
where $\gamma^\mathscr{B'}$ is the corresponding anomalous dimension. The solution of the evolution equation (\ref{anomalyb}) is
\begin{eqnarray}
\mathcal{C}^{\mathscr{B'}} _j (E,u,\mu)&=&\left(\frac{2E}{\mu_Q}\right)^{a(\mu_Q ,\mu)}e^{S(\mu_Q,\mu)} \int_0 ^1 dv\  U_{\perp,\parallel}(u,v,\mu_Q , \mu)\mathcal{C}_j ^{\mathscr{B'}}(E,v,\mu_Q). \label{CBmu}
\end{eqnarray}
Here prime notation is adopted to make it consistent with our earlier tree level coefficients in Eq. (\ref{tlevelC}). The functions $a(\mu_Q ,\mu)$ and $S(\mu_Q ,\mu)$ appeared in Eq. (\ref{CBmu}) are defined as \cite{Bosch:2003fc}
\begin{eqnarray}
S(\mu_Q , \mu) &=& \frac{\Gamma_0}{4\beta_0 ^2}\left[\frac{4\pi}{\alpha_s (\mu_Q)}\left(1-\frac{1}{r_1}-\text{ln} r_1\right)+\frac{\beta_1}{2\beta_0}\text{ln}^2 r_1 -\left(\frac{\Gamma_1}{\Gamma_0}-\frac{\beta_1}{\beta_0}(r_1 -1-\text{ln} r_1)\right)\right]\nonumber\\
a(\mu_Q ,\mu)&=&-\frac{\Gamma_0}{2\beta_0}\text{ln} r_1,
\end{eqnarray}
where $\Gamma_0 =4C_F$ and
\begin{eqnarray} 
\Gamma_1 &= &4C_F \left[\left(\frac{67}{9}-\frac{\pi^2}{3}\right)C_A -\frac{20}{9}T_F n_f\right], \notag\\
\beta_0 &=&\frac{11}{3}C_A -\frac{4}{3}T_F n_F\notag\\
\beta_1 &=& \frac{34}{3}C_A ^2 - \frac{20}{3}C_A T_F n_f -4C_F T_F n_f, \notag\\
r_1&=&\alpha_s(\mu)/\alpha_s(\mu_Q).
\end{eqnarray}
%$\tilde{\gamma}_0=-5C_F$,
To find the evolution kernels $U_{\perp,\parallel}$, we employed the method used in \cite{Hill:2004if} to solve the RG equation. Using the  initial condition $U_{\perp,\parallel}(u,v,\mu_Q ,\mu_Q)=\delta(u-v)$, we have
\begin{eqnarray}
\frac{d}{d \text{ln} \mu}\frac{U_{\perp,\parallel}(u,v,\mu_Q ,\mu )}{\bar{u}}&=& \int_0 ^1 dy y \bar{y}^2 \frac{V_{\perp,\parallel}}{\bar{y}\bar{u}}\frac{U_{\perp,\parallel}(u,v,\mu_Q ,\mu)}{\bar{y}}+w(u)\frac{U_{\perp,\parallel}(u,v,\mu_Q ,\mu )}{\bar{u}}.\nonumber\\\label{evolU}
\end{eqnarray}
\begin{figure}
\centering
  \includegraphics[width=100mm]{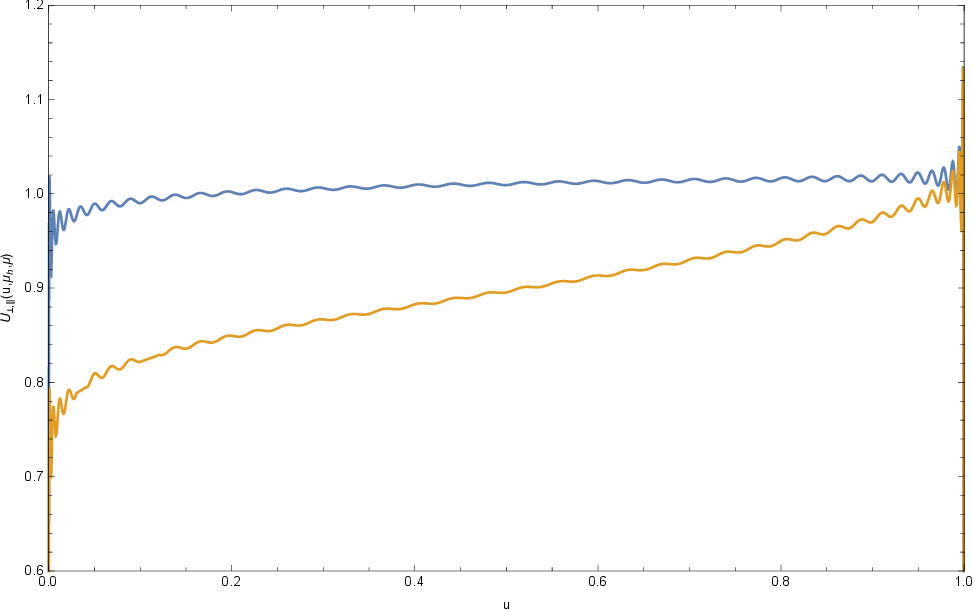}
  \caption{Evolution function with blue being $U_\parallel$ and orange being $U_\perp$ using 80 basis functions.}
  \label{evolution}
\end{figure}
The functions $V_{\perp,\parallel}$ and $w(u)$ are defined in \cite{Hill:2004if} which are  the anomalous dimension contributions. A basis function comprising of Jacobi Polynomials (obtained by solving eigenvalue equation for $V_{\perp,\parallel}$) is used to construct the solution for Eq. (\ref{evolU}). This solution is plotted in Fig. \ref{evolution} by taking a set of 80 basis functions. In the case of radiative decays, the perpendicular evolution function $U_{\perp}$ is important. \footnote{In our previous paper \cite{Sikandar:2021shv}, the perpendicular and parallel components were mistakenly interchanged in the text.}

\section{Matrix elements and Branching Ratios}\label{sec6}
%-------------------------------------------------------

%In $B \to T$ decays, where $T$ representing $K_2^*,\; a_2$ or $f_2$ meson, is taken to be light compared to the initial state $B$-meson. However, in contrast to the light mesons $(\pi, \rho)$, the masses of these tensor mesons are above the hadronic scale, therefore a special care is warranted while studying the perturbative and non-perturbative parts in $B \to T$ decays. It will allow us to use special factorization scheme, we have introduced in Section  \ref{SBCs}.  
 Let $p_B^{\mu}(p_T^{\mu})$ and $m_B(m_T)$ are the momentum and mass of $B(T)$ meson, respectively. The polarization of the spin-2 meson can be written as $\varepsilon_{\mu\nu}(\lambda)$, with $\lambda$ representing the helicity. The polarization tensor satisfying  $\varepsilon_{\mu\nu}(\lambda)p_T^\nu=0$ is traceless and symmetric. Hence, it is easy to construct it from the polarization states of a massive vector (spin-1) particle, i.e., 
\begin{eqnarray}
\varepsilon_{\mu\nu}(\pm 2)&=&\varepsilon(\pm)_\mu\varepsilon(\pm)_\nu,\notag\\
\varepsilon_{\mu\nu}(\pm 1)&=&\sqrt{\frac{1}{2}}\left(\varepsilon(\pm)_\mu \varepsilon(0)_\nu +\varepsilon(0)_\mu\varepsilon(\pm)_\nu\right), \notag\\
\varepsilon_{\mu\nu}(\pm 0)&=&\sqrt{\frac{1}{6}}\left(\varepsilon(+)_\mu \varepsilon(-)_\nu +\varepsilon(-)_\mu\varepsilon(+)_\nu\right)+\sqrt{\frac{2}{3}}\varepsilon(0)_\mu\varepsilon(0)_\nu.
\end{eqnarray}
Here, the massive vector states $\varepsilon_\mu(\pm)$ and $\varepsilon_\mu(0)$ are given as;
\begin{eqnarray}
\varepsilon_\mu(0)&=&\left(\frac{|\vec{p}_T|}{m_T},0,0,\frac{E_F}{m_T}\right),\notag\\
\varepsilon_\mu(\pm)&=&\frac{1}{\sqrt{2}}(0,\mp 1,-i,0).
\end{eqnarray}
Here
$E_F=\frac{m_B ^2 +m_{T}^2}{2m_B}$ 
is the energy of the final state tensor meson. At the maximum recoil, its value is
 $E_F\sim E= m_B/2$, while its 3-momentum is,  $|\vec{p}_T|=\sqrt{E_F^2-m_B^2}$.  It is useful to introduce a new polarization vector $\varepsilon_{\perp \mu}$
\begin{eqnarray}
\varepsilon_{\perp\mu}(\lambda)&=&\varepsilon_{\mu\nu}(\lambda)\frac{p_T^\nu}{m_B},\label{polvec}
\end{eqnarray}
such that $\varepsilon_{\perp\mu}(\pm 2)=0 ,\quad \varepsilon_{\perp\mu}(\pm 1)=\left(|\vec{p}_T|/m_B\right)\beta_\perp\varepsilon_\mu(\pm) ,\quad\varepsilon_{\perp\mu}(\pm 0)=(|\vec{p}_T|/m_B)\alpha_L\varepsilon_\mu(0)$ with $\beta_\perp =\sqrt{1/2}$, and $\alpha_L =\sqrt{2/3}$.  
Using the transversality condition $\varepsilon_\perp^*\cdot p_T =0$ gives
\begin{eqnarray}
\varepsilon^*_\perp \cdot n =-\frac{m_{_{T}}^2}{4E|\vec{p}_T|}\varepsilon^*_\perp\cdot \bar{n}.\label{identity}
\end{eqnarray}
The photon has polarization $\epsilon^*$ and for the real photon $\epsilon^*\cdot p_\gamma=\epsilon^*\cdot \bar{n} =0$. The $B\rightarrow T$ form factors can be parameterized in terms of soft-overlap function $\zeta^\perp_{T}$ by using SCET$_{\text{I}}$ $\mathscr{A}-$type operator
\begin{eqnarray}
\langle T(p_T) |\bar{\chi}_{_{hc}}\Gamma h| B(v)\rangle &=-2E_F \zeta^\perp_{T}(E_F)\text{tr}[\bar{\mathcal{M}}_{T\perp}(n)\Gamma\mathcal{M}_B (v)],\label{matrixelements}
\end{eqnarray}
where the projection operators, $\mathcal{M}_{T\perp}(n)$ and $\mathcal{M}_B (v)$ read as
\begin{eqnarray}
\mathcal{M}_B (v)&=-\frac{1+\slashed{v}}{2}\gamma_5,\qquad \mathcal{\overline{M}}_{T \perp} (n)=\slashed{\varepsilon}_\perp ^* \frac{\slashed{\bar{n}}\slashed{n}}{4}.
\label{proj}
\end{eqnarray}
\begin{table}[t]
  \centering
    \begin{tabular}{|l|l|l|l}
    \hline
    \multicolumn{3}{|c|}{\textbf{Input Values}} \\ \hline
     $m_B$\qquad\qquad\qquad\quad  5.28 GeV  & $m_{b}(\bar{m}_b)$\cite{FlavourLatticeAveragingGroup:2019iem}\qquad\qquad $4.18$ GeV &  $\tau_B$\cite{Workman:2022ynf} \qquad\qquad 1.519 ps    \\ \hline

   $m_{{K}_2^*}$\qquad\qquad\qquad\ 1.43 GeV   &$m_{{f}_2}$\qquad\qquad\qquad\quad  1.27 GeV& $m_{{a}_2}$\qquad\qquad\qquad\  1.32 GeV \\ \hline

        $f_{K_2^*}^{\perp}$(1GeV)\cite{Yang:2010qd}\quad 0.077$\pm$ 0.014 GeV  &  $f_{f_2}^\perp$(1GeV)\cite{Yang:2010qd}\quad 0.117$\pm$.025 GeV&$f_{a_2}^\perp$(1GeV)\cite{Yang:2010qd}\quad 0.105$\pm$0.021 GeV\\ \hline

 $f_{B}$\cite{FlavourLatticeAveragingGroup:2019iem}\qquad\qquad\qquad 0.190 GeV & $G_F$\qquad\qquad $1.16 \times 10^{-6}$ GeV &$|V^*_{cd}V_{cs}|$\qquad\qquad $0.0088$ \\ \hline

  $|V^* _{cs}V_{cb}|$\qquad\qquad\qquad\ \  $4\times 10^{-2}$   & $a_{1\perp}^{T}$(1GeV)\qquad\qquad\quad 5/3&$a_{1\parallel}^{T}$(1GeV)\qquad\qquad\quad 5/3\\ \hline

      \end{tabular}
    \caption{The values of input parameters used in the numerical analysis.}\label{TABLE2}
\end{table}
%Substituting these projectors from Eq. (\ref{proj}) in Eq. (\ref{matrixelements}) with $\Gamma = (1+\gamma_5)$, we get
%\begin{eqnarray*}
%\langle T(p_T,\varepsilon_\perp ^*)\gamma(\kappa^*) |J^\mathscr{A}| B(v)\rangle =\mathcal{C}^\mathscr{A} E_F \zeta_{T\perp}(E_F)\left[2 (\varepsilon^*_\perp\cdot \kappa^*)-(n\cdot\varepsilon^*_\perp)(\bar{n}\cdot\kappa)\right. \\
%\left. -(n\cdot\kappa^*)(\bar{n}\cdot\varepsilon^*_\perp)+i\epsilon^{\alpha\beta\mu\nu}n_\alpha\bar{n}_{\beta}\varepsilon^*_{\perp\mu}\kappa^*_\nu\right].
%\end{eqnarray*}
%As $\bar{n}\cdot\varepsilon^*=0$ and because both photon and axial-vector meson are left circularly polarized, $\varepsilon^* \cdot \eta^* =1$. Making use of Eq. (\ref{identity}) and $\epsilon^{0123} =-1$, we get $\epsilon^{\alpha\beta\mu\nu}n_\alpha\bar{n}_{\beta}\varepsilon^{*}_\mu\eta^*_\nu =-2i$. This leads to
This results in
\begin{eqnarray}
\langle T \gamma |J^\mathscr{A}| B\rangle =4 \mathcal{C}^{\mathscr{A}}E_F\left(1+\frac{m^2_{T}}{8E_F|\vec{p}_T|}\right)\zeta^\perp_{T}(E_F).\label{Aresult}
\end{eqnarray}
The matrix elements of $\mathscr{B}^\prime-$type currents $O_{1,2}^\mathscr{B^\prime}$ of SCET$_{\text{II}}$, with independent soft and collinear parts can be written as a convolution of the LCDA's of the respective meson. We can write it as 
\begin{eqnarray}
\langle 0| \bar{\mathcal{Q}}_s(tn)\frac{\slashed{n}}{2}\Gamma \mathcal{H}(0)|\bar{B}(v)\rangle &=&-\frac{i\sqrt{m_B}F(\mu)}{2}\text{Tr}\left(\frac{\slashed{n}}{2}\Gamma\frac{1+\slashed{v}}{2}\gamma_5 \right)\times\int_0 ^\infty d\omega e^{-i\omega t n\cdot v }\phi_B (\omega, \mu,\nonumber\\)\label{blcda}\\
\langle T_{\perp}(p_F,\eta^*)|\bar{\chi}_c (s_1\bar{n})\Gamma \frac{\bar{n}}{2}\chi_c (0)|0\rangle&=&-\frac{if_{T_{\perp}}(\mu)}{4}\bar{n}\cdot p_T \text{Tr}\left(\slashed{\varepsilon}^*_\perp \Gamma \frac{\slashed{\bar{n}}\slashed{n}}{4}\right)\times\int_0 ^\infty du e^{ius_1\bar{n}\cdot p_T }\phi_{T_{\perp}}(u,\mu),\nonumber\\\label{klcda}
\end{eqnarray}
where $\phi_B$ and  $\phi_{T\perp}$ are distribution amplitudes for $B$ and $T$ mesons, respectively. $f_{T\perp}$ is the decay constant of a tensor meson and it depends upon the scale of the theory. The scale dependent quantity $F(\mu)$ is related to $B-$meson decay constant $f_B$ at NLO as
\begin{eqnarray*}
f_B\sqrt{m_B}&=F(\mu)\left(1+\frac{C_F \alpha_s (\mu)}{4\pi}\left(3\text{ln}\left(\frac{m_b}{\mu}\right)-2\right)\right).
\end{eqnarray*}
Collecting the results from Eqs. (\ref{Aresult}, \ref{blcda}) and Eq. (\ref{klcda}) to get the resummed matrix elements at leading power and at an order $\mathcal{O}(\alpha_s)$, we have
\begin{eqnarray}
\mathcal{M}=\mathcal{C}^\mathscr{A}(E_F,\mu)\zeta^\perp_{T}(E_F)&-&\frac{\sqrt{m_B}}{4}\left(\frac{2E_F}{\mu_Q}\right)^{a(\mu_Q ,\mu_i)}e^{S(\mu_Q,\mu_i)}\int_0 ^\infty \frac{d\omega}{\omega}\phi_B (\omega, \mu)
 \int_0 ^1 du  f_{T\perp}\phi_{T_{\perp}}(u,v)\nonumber\\
&\times&\int_0^1dv \mathcal{J}_\perp \left(u,v,\text{ln}\frac{m_B \omega}{\mu^2}\right)U_{\perp}(u,v,\mu_Q , \mu_i)\mathcal{C}_1 ^\mathscr{B'}(v,\mu).\label{matrixel}
\end{eqnarray}
The complete Wilson coefficients $C^{\mathscr{A}}$  and $C^{\mathscr{B}'}$ are given as
\begin{eqnarray}
C^{\mathscr{A}}&=& \frac{G_F}{\sqrt{2}}V^*_{cp}V_{cb}\left[C_7+\frac{C_F\alpha_s}{4\pi}(C_1G_1(x_c)+C_8G_8)\right] \nabla_7C^{\mathscr{A}},\\
C_1^{\mathscr{B}'}&=& \frac{G_F}{\sqrt{2}}V^*_{cp}V_{cb}\left[C_7+C_1\frac{1}{3}f\left(\frac{\bar{m}_c^2}{4\bar u E E_\gamma}\right)+C_8\frac{\bar{u}}{3u}\right] \nabla_7C^{\mathscr{B}'},
\end{eqnarray}
where $p=s$ and $d$ for $K_2^*$ and $(f_2,a_2)$ mesons, respectively. The function $G_1(x_c)$ and $G_8$ are given in Appendix \ref{App}.
The LCDA for the tensor meson is defined as \cite{Yang:2010qd}
\begin{eqnarray}
\phi_{T\perp}&=&6u\bar{u}\left[1+3a_1^\perp \Upsilon +\frac{3}{2}a_2^\perp (5\Upsilon^2-1)\right],
\end{eqnarray}
with $\Upsilon=2u-1$ and it is normalized as $\int_0^1\phi_\perp (u)/u=1$. The Gegenbaur moments and decay constants are given in the Table \ref{TABLE2}. The distribution amplitudes for $B-$meson is a non-perturbative input heavily prone to uncertainty. Recent theoretical studies \cite{Beneke:2011nf,Wang:2016qii} constrain the inverse $B-$moment (defined as $\lambda_B^{-1}=\int_0^\infty d\omega\ \Phi_{B}^{+}/\omega$). In literature, \cite{Wang:2015vgv, Gao:2019lta} the running of $\lambda_B^{-1}$ is discussed.  It is pertinent to mention that in principle one has to employ the RG evolution in the matching of SCET$_{\text{I}}$ to SCET$_{\text{II}}$ operators, and to do so one has to solve the one-loop SCET$_{\text{II}}$ diagrams to find the corresponding anomalous dimension. But, due to the close values of intermediate scale $\mu_i \sim 1.5$ GeV and the hadronic scale, this is not required, and the values of decay constants and inverse $B$ moment are taken at fixed value, i.e., 1 GeV  just like the one in  \cite{Ali:2006ew}. Therefore, here we took an optimal value of $\lambda_B^{-1}=0.35\pm 0.10$GeV$^{-1}$. The difference of the optimal value of  $\lambda_B^+(1\text{GeV})$ we used here and the one given in  \cite{Ali:2006ew} is due to the constraints discussed by Beneke \textit{et. al.} \cite{Beneke:2011nf}.

The soft overlap function, $\zeta^\perp_{T}$, which is also the non-factorizable part and required as a nonperturbative input, is estimated from the form-factor values calculated in \cite{Aliev:2019ojc} by using Light-cone sum rules (LCSR) . The seven form-factors for $B\rightarrow T$ can be related to $\zeta_{T}^\perp$ by invoking the heavy quark symmetry, therefore, to determine $\zeta^\perp_{T}$ at zero momentum transfer i.e., $q^2=0$, any one of them can be used \cite{Sikandar:2022iqc}. Using this liberty, let us use the vector form-factor $V(q^2)$, which is calculated by using LCSR \cite{Wang:2010ni,Aliev:2019ojc}, to determine the $\zeta^\perp_{T}$, i.e., 
\begin{equation}
V(q^2)=\left(1+\frac{m_T}{m_B}\right)\frac{E_F}{|\vec{p}_T|}\mathcal{C}^{\mathscr{A}}_{V1}\zeta^\perp_{T}(E_F).
\end{equation}
Here the Wilson Coefficient corresponding to a vector current at NLO is given as
\begin{eqnarray}
C_{V1}^{\mathscr{A}}&=&1-\frac{\alpha_s C_F}{4\pi}\left[\left(\frac{1}{1-x}-3\right)\text{ln}(x)+2\text{ln}^2(x)+2\text{ln}^2(\frac{\mu_Q}{m_b})^2+2\text{Li}_2(1-x)-(4\text{ln}(x)-5)\text{ln}(\frac{\mu_Q}{m_b})+\frac{\pi^2}{12}+6\right]\nonumber\\
\end{eqnarray}
The soft-overlap functions are thus calculated to be
\begin{align}
\zeta^ \perp_{K_2^*}(0)&=0.295^{+0.056}_{-0.056},\notag\\
\zeta^ \perp_{a_2 }(0)&=0.288^{+0.06}_{-0.05},\notag\\
\zeta^ \perp_{f_2 }(0)&=0.185^{+0.12}_{-0.08}. \label{softoverlap}
\end{align}
The theoretical errors in the branching ratio of $B\rightarrow f_2 \gamma$ compared to the other two decays are higher due to the higher uncertainty in the value of the form-factor calculated in \cite{Aliev:2019ojc}. For consistency check, we also determined these soft overlap functions from the tensor current having an $\alpha_s$ contribution, and found that they lie within 5\% of the values found in Eq. (\ref{softoverlap}). Comparing this value of the $K_2^*$ with the corresponding overlap function for the $K^{*}$ $\zeta^\perp _{K^*}=0.37\pm0.04$ estimated from $B \to K^*\gamma$ \cite{Becher:2005fg} and $\zeta^\perp_{K^*}(q^2=0)=0.32\pm 0.02$ calculated from the semileptonic $B \to K^*$ decay \cite{Ali:2006ew}, our result for $B\to K_2^*$ soft form-factor has comparatively large uncertainty coming from the LCSR calculation of the form-factor $V(q^2 = 0)$.
The RG-improved branching ratios for $B\rightarrow \left(K_2^*,f_2,a_2\right)\gamma$ decays at NLO is given as 
\begin{equation}
\mathscr{B}(B\rightarrow (K_2^*,f_2,a_2)\gamma)=\frac{\tau_B m_B}{4\pi}\left(1+\frac{m_T^2}{8E_F|\vec{p}_T|}\right)\left(1-\frac{m_T^2}{m_B^2}\right)|\mathcal{M}|^2,
\end{equation}
Using these overlap functions, the branching ratios for $B\to \left(K_2^*,\; f_2,\; a_2\right)\gamma$ decays are calculated and these are given in Table (\ref{TABLE3}). The first associated uncertainty is due to soft-overlap function while the second is the hadronic uncertainty coming from inverse $B-$moment and the decay constants. The dominant contribution to the branching fraction is due to the first term in the amplitude Eq. (\ref{matrixel}), which we can call as the soft-amplitude $(\mathscr{M}_{\text{soft}})$ while the second term in (\ref{matrixel}) may be termed as hard-amplitude $(\mathscr{M}_{\text{hard}})$. The contribution of $\mathscr{M}_{\text{hard}}$ to the total branching ratio of $B\rightarrow K_2^*\gamma$ is 26\% of the value of the lone contribution from $\mathscr{M}_{\text{soft}}$. The hard-amplitude is itself RG-improved and induces 6.5\% correction to the branching ratio compared to the one without RG-improvement. For $B\to K_2^{*}\gamma$, our result of the branching ratio is compared with the values calculated using LFQM \cite{Cheng:2003sm,Cheng:2009ms}, HQET \cite{Veseli:1995bt} and the experimental measured averaged value \cite{Workman:2022ynf}. Our result (although prone to uncertainty) is closer to the experimental average. The branching ratio is very sensitive to the soft-overlap function and its precise measurement reduces the uncertainties significantly.
\begin{table}[t]
  %\centering
    \begin{tabular}{|c|c|c|c|c|c|}
    \hline
    \multicolumn{5}{|c|} {Branching ratios with a factor of 10$^{-6}$ for $B\to (K_2^{*},\;f_2,\; a_2)\gamma$ decays.} \\ \hline
    Decay mode&(this work )& LFQM \cite{Cheng:2003sm,Cheng:2009ms} & HQET\cite{Veseli:1995bt}&Experimental average\cite{Workman:2022ynf}\\ \hline
    $B\rightarrow K_2^*(1430)\gamma$ & $16.7^{+6.36}_{-6.36} \pm 1.2$&$29.4^{+31.8}_{-13.9}$&$21.8\pm10.2$&$12.4\pm 2.4$\\ \hline
    $B\rightarrow a_2(1320)\gamma $&$0.67^{+0.30}_{-0.23}\pm 0.06$&$-$&$-$&$-$\\ \hline
    $B\rightarrow f_2(1270)\gamma$&$0.18^{+0.23}_{-0.16}\pm 0.08$&$-$&$-$&$-$\\ \hline
        \end{tabular}
    \caption{The results of branching ratios of $B\to (K_2^{*},\; a_2,\; f_2)\gamma$ decays calculated in this work. In the case of $B\to K_2^*\gamma$, it is compared with the corresponding PDG averaged value, and with the values calculated using LFQM and HQET. }\label{TABLE3}
\end{table}

\section{Conclusion}\label{sec7}
The rare radiative $B\to K^*$ decays were studied using HQET/LEET and SCET \cite{Beneke:2003pa,Becher:2003qh}. By changing the final state vector mesons with the corresponding axial-vector meson $\left(J^P=1^+\right)$, the branching ratios of $B\rightarrow (K_1,b_1,a_1)\gamma$ decays in the SECT are found in \cite{Sikandar:2021shv}. In this work, we employed SCET to calculate the branching ratio of the radiative $B$ to tensor meson decays, i.e., $B\rightarrow (K_2^*,f_2,a_2)\gamma$ at NLO in strong coupling constant $\left(\alpha_s\right)$ and at leading power of $1/m_b$. The QCD diagrams matched with SCET$_\text{I}$ at the loop (tree) level for $\mathscr{A}(\mathscr{B}^\prime)-$ type operators. For the $\mathscr{B^\prime}-$ type operators, we resummed the large logarithms using the one loop diagrams for the SCET$_\text{I}$. Using the value of soft form factor (soft-overlap function) estimated using LCSR approach  \cite{Aliev:2019ojc}, we calculated the branching ratios for $B\rightarrow (K_2^*,f_2,a_2)\gamma$ decays. We found that our result (central value) of the branching ratio of $B\rightarrow K_2^* \gamma$ are smaller than the corresponding LFQM \cite{Cheng:2003sm,Cheng:2009ms} and HQET \cite{Veseli:1995bt}, but closer to the PDG experimental average \cite{Workman:2022ynf}. The other important feature of the calculation done here is that the the errors in the branching ratio of $B\rightarrow K_2^* \gamma$ are small compared to \cite{Cheng:2003sm,Cheng:2009ms, Veseli:1995bt}, because we used LCSR value of $V(q^2 = 0)$ having small uncertainties to extract soft-overlap function $\left(\zeta_T^{\perp}\right)$. In contrast with  $B\rightarrow K_2^* \gamma$, the branching ratios of $B\to \left(a_2, f_2\right)\gamma$ are CKM suppressed, and we hope that these would be measured in some ongoing and future $B$ physics experiments. 
\section{Appendix}\label{App}

For the matching of $Q_1$ and $Q_8$ on $\mathscr{A}-$type operators, the functions $G_1$ and $G_8$ were required. These are taken from \cite{Bosch:2001gv}, and their explicit form is
\begin{eqnarray}
G_1(x)&=&\frac{-104}{27}\text{ln}\frac{\mu }{m_b}-\frac{833}{162}-\frac{20 \pi i}{27}+8\frac{8\pi^2}{9}x^{3/2} \nonumber\\
&+&\frac{2}{9}\left[48 +30 \pi i-5\pi^2-36\xi(3)+(36+6\pi i-9\pi^2)\text{ln}(x) +(3+6\pi i)\left(\text{ln}(x)\right)^2 +\left(\text{ln}(x)\right)^3\right]x\nonumber\\
&+&\frac{2}{9}\left[18 +2\pi^2 -2\pi^3 +(12-6\pi^2)\text{ln}x +6\pi i \left(\text{ln}(x)\right)^2+\left(\text{ln}(x)\right)^3 \right]x^2\nonumber\\
&+&\frac{1}{27}\left[-9+112\pi i-14\pi^2+(182-48\pi i)\text{ln}(x) -126\left(\text{ln}(x)\right)^2 \right]x^3 +\mathcal{O}(x^4),\nonumber\\
G_8&=& \frac{8}{3}\text{ln} \left(\frac{\mu }{m_b}\right)+\frac{11}{3}+\frac{2\pi i}{3}-\frac{2\pi^2}{9},
\end{eqnarray}
here $x=m_c^2/m_b^2$. The function $f(z)$, where $z=\frac{m_q^2}{4EE_\gamma \bar{u}}$, given in the matching of $Q_1$ operators to $\mathscr{B^\prime}$ and $\mathscr{C}-$types currents is
\begin{eqnarray}
f(z)=\left\{
  \begin{array}{@{}ll@{}}
    1+4z\left(\text{arctanh}(\sqrt{1-4z})-i\frac{\pi}{2}\right)^2, & \text{for}\ z<1/4 \\\\

    1-4z\ \text{arctan}^2\frac{1}{\sqrt{4z-1}}, & \text{for}\ z>1/4.
  \end{array}\right.
 \end{eqnarray}

%\section*{Acknowledgments}

\restoregeometry

\end{document}